\documentclass[12pt]{article}
\usepackage{amssymb,amsfonts,amsmath}
\begin{document}

\baselineskip=.22in
\renewcommand{\baselinestretch}{1.2}
\renewcommand{\theequation}{\thesection.\arabic{equation}}
~\vspace{12mm}

\begin{center}
{{{\Large \bf Vortex-type Half-BPS Solitons in ABJM Theory}
}\\[12mm]
{Chanju Kim}\\[3mm]
{\it Institute for the Early Universe and Department of Physics\\
Ewha Womans University,
Seoul 120-750, Korea}\\
{\it and}\\
{\it Korea Institute for Advanced Study, 130-722, Korea}\\
{\tt cjkim@ewha.ac.kr}\\[6mm]
{Yoonbai Kim,~~O-Kab Kwon,~~Hiroaki Nakajima}\\[3mm]
{\it Department of Physics, BK21 Physics Research Division,
and Institute of Basic Science\\
Sungkyunkwan University, Suwon 440-746, Korea}\\
{\tt yoonbai, okab, nakajima@skku.edu}
}
\end{center}
\vspace{10mm}

\begin{abstract}
We study Aharony-Bergman-Jafferis-Maldacena (ABJM) theory 
without and with mass 
deformation. It is shown that maximally supersymmetry preserving, 
D-term, and F-term mass deformations of single mass parameter are equivalent.
We obtain vortex-type half-BPS equations and the corresponding energy bound. 
For the undeformed ABJM theory, the resulting half-BPS
equation is the same as that in supersymmetric Yang-Mills theory and
no finite energy regular BPS solution is found.
For the mass-deformed ABJM theory, the half-BPS equations for U(2)$\times$U(2) 
case reduce to the vortex equation in Maxwell-Higgs theory, which supports
static regular multi-vortex solutions.  
In U($N$)$\times$U($N$)
case with $N>2$ 
the nonabelian vortex equation of Yang-Mills-Higgs theory 
is obtained. 
\end{abstract}


\newpage

\setcounter{equation}{0}
\section{Introduction}\label{sec1}

The low-energy limit of M-theory is 11-dimensional supergravity 
and involves membranes and five-branes as solitonic
extended objects~\cite{Townsend:1996xj}. 
Recently, the Bagger-Lambert-Gustavsson (BLG) 
theory~\cite{Bagger:2006sk,Gustavsson:2007vu} and 
the Aharony-Bergman-Jafferis-Maldacena (ABJM) theory~\cite{Aharony:2008ug}
have been proposed as the low-energy limit of world-volume theory of multiple 
M2-branes. The ABJM theory is given in the basis of brane constructions   
and is described by (1+2)-dimensional Chern-Simons-matter 
theories with U($N$)$\times$U($N$) or SU($N$)$\times$SU($N$) gauge
group and ${\cal N}$=6 supersymmetry (SUSY). 
In large $N$ limit, the ABJM theory
is dual to M-theory on AdS${}_{4}\times {\rm S}^{7}/{\mathbb Z}_{k}$,
where $k$ is related with the discrete level of Chern-Simons term.

When the world-volume theory of the stacked M2-branes is constructed, 
one of the main tests is to reproduce Basu-Harvey fuzzy 
funnel~\cite{Basu:2004ed} as a BPS configuration~\cite{Bagger:2006sk}. 
Along this line, the composite of M2-M5 and the domain 
wall solutions are studied in the BLG 
theory~\cite{Hosomichi:2008qk,Krishnan:2008zm,Bonelli:2008kh}
and the ABJM theory~\cite{Terashima:2008sy} without and with 
mass deformation. When these codimension-one BPS objects are dealt,
role of the two Chern-Simons gauge fields is completely missing.  
Among diverse research directions in the world-volume theory of M2-branes, 
it deserves to investigate BPS solitons for which the gauge fields play
a crucial role. These are nothing but point-like BPS Chern-Simons vortices.

Relativistic Chern-Simons-Higgs theory with sextic scalar potential  
is first introduced in order to saturate the BPS bound 
for the static multi-BPS vortex solutions~\cite{Hong:1990yh}, and
also arises in the supersymmetric abelian Chern-Simons-Higgs
theories~\cite{Lee:1990it}.
An attractive point of the BPS limit of Chern-Simons-Higgs theories
is rich soliton spectrum due to coexistence of both the symmetric and
broken vacua. In addition to the topologically stable multi-BPS vortices
and domain walls, 
marginally stable nontopological solitons (or Q-balls) and 
nontopological vortices (or Q-vortices) 
exist~\cite{Jackiw:1990pr}.  
Extension to U(1)$\times$U(1) gauge group~\cite{Kim:1992uw} and nonabelian 
gauge group~\cite{Cugliandolo:1990nb} is also made.
Therefore, in the scheme of (1+2)-dimensional quantum field theories, 
the mass-deformed BLG and ABJM theories are understood as the 
complicated Chern-Simons-Higgs theories, 
and the undeformed BLG and ABJM theories as their superconformal limit.
 
In this paper, we first discuss relation among the proposed mass deformations 
in the ABJM theory, with single mass deformation parameter. One is
maximally supersymmetric mass deformation in terms of ${\cal N}=1$ superfield
formalism~\cite{Hosomichi:2008qk} and in component fields~\cite{Gomis:2008vc}. 
Two other types of mass deformation correspond to D-term and F-term 
deformations in the basis of ${\cal N}=2$ superfield 
formalism~\cite{Gomis:2008vc}. Though they look different theories possessing 
different manifest SUSY's, we shall show that the aforementioned 
three mass deformations 
are equivalent.
The main subject of our interest is to understand static vortex-type 
half-BPS objects of the ABJM theory without and with mass deformation.
In BLG theory of SU(2)$\times$SU(2) gauge symmetry, possible BPS equations 
were classified and some vortex-type BPS configurations were
obtained~\cite{Jeon:2008bx,Jeon:2008zj,Kim:2008cp}.  
In ABJM theory, some Chern-Simons vortex-type $\frac{1}{6}$-BPS solitons were
obtained, including topological vortices, nontopological solitons, and
nontopological vortices~\cite{Arai:2008kv}, and vortex loop operators were
constructed~\cite{Drukker:2008jm}. Here we examine half-BPS equations for
static vortex-type solitons in the U($N$)$\times$U($N$) ABJM theory
both without and with mass deformation, and discuss in 
detail the possible singular and regular multi-BPS vortex solutions.
Though various point-like solitons are obtained in the
world-volume theory of M2-branes as singular solutions without mass 
deformation and regular solutions with mass deformation, they await
proper interpretation in the context of M-theory.

The rest of this paper is organized as follows. We begin section~\ref{sec2}
with introduction of the ABJM theory, and in subsection~\ref{ss21}
we discuss the relation among three proposed mass deformations. 
In section~\ref{sec3} vortex-type half BPS equations 
and the corresponding energy bound 
are obtained. In section~\ref{sec4} we reduce the 
general half-BPS equations in undeformed theory 
in a simple set of two coupled first-order 
equations which is the same as that in supersymmetric Yang-Mills theory with
the Yang-Mills coupling identified as what obtained in D2 
limit of the theory~\cite{Mukhi:2008ux,Pang:2008hw}.
There is no static vortex-like half-BPS solution with finite energy.
In section~\ref{sec5}, we examine half-BPS equations in mass-deformed theory.
We first consider U(2)$\times$U(2) case, leading to the vortex equation in 
Maxwell-Higgs theory, and find spinless multi-BPS vortex
vortices without or with constant background magnetic field.
For U($N$)$\times$U($N$) case with $N>2$, under a suitable ansatz,
the BPS equations reduce to the nonabelian vortex equation in Yang-Mills-Higgs
theory. We also obtain other equations with different ansatz.
We conclude in section~\ref{sec6} with brief summary and discussion.

\setcounter{equation}{0}
\section{ABJM Theory with and without Mass Deformation}\label{sec2}

The ABJM theory is an ${\cal N}=6$ superconformal 
${\rm U}(N)\times {\rm U}(N)$
Chern-Simons theory with level $(k,-k)$ coupled to four complex scalars
and four Dirac fermions in the bifundamental representation,
\begin{align}\label{ABJMa}
S_{{\rm ABJM}} =  \int d^3x\: \bigg\{
    &\frac{k}{4\pi} \epsilon^{\mu\nu\lambda} {\rm tr}\left(
           A_\mu \partial_\nu A_\lambda + \frac{2i}{3} A_\mu A_\nu A_\lambda
         - \hat{A}_\mu \partial_\nu \hat{A}_\lambda - \frac{2i}{3}
         \hat{A}_\mu \hat{A}_\nu \hat{A}_\lambda\right)
\nonumber \\
& - {\rm tr} \big(D_\mu Y_{A}^\dagger D^\mu Y^{A} \big)
    + {\rm tr}\big( \psi^{A\dagger} i\gamma^{\mu}D_{\mu} \psi_{A}\big)
    - V_{{\rm ferm}} - V_0 \bigg\},
\end{align}
where $A = 1,\ldots,4$ and
\begin{equation}
D_{\mu}Y^A = \partial_{\mu} Y^A + iA_{\mu}Y^A-iY^A{\hat A}_{\mu}.
\end{equation}
$V_{{\rm ferm}}$ is the Yukawa-type quartic-interaction term,
\begin{align}
V_{{\rm ferm}}=\frac{2i\pi}{k}\, {\rm tr}\big(
&Y_{A}^{\dagger}Y^{A}\psi^{B\dagger}\psi_{B}
-Y^{A}Y_{A}^{\dagger}\psi_{B}\psi^{B\dagger}
+2Y^{A}Y_{B}^{\dagger}\psi_{A}\psi^{B\dagger}
-2Y_{A}^{\dagger}Y^{B}\psi^{A\dagger}\psi_{B} \nonumber\\
& -\epsilon^{ABCD}Y_{A}^{\dagger}\psi_{B}Y_{C}^{\dagger}\psi_{D}
+\epsilon_{ABCD}Y^{A}\psi^{B\dagger}Y^{C}\psi^{D\dagger}
\big),
\end{align}
and $V_0$ is the sextic scalar potential,
\begin{align} \label{potential0}
V_0= -\frac{4\pi^{2}}{3k^{2}}\, {\rm tr}\big(&
Y^{A}Y_{A}^{\dagger}Y^{B}Y_{B}^{\dagger}Y^{C}Y_{C}^{\dagger}
+Y_{A}^{\dagger}Y^{A}Y_{B}^{\dagger}Y^{B}Y_{C}^{\dagger}Y^{C} \nonumber\\
& +4Y^{A}Y_{B}^{\dagger}Y^{C}Y_{A}^{\dagger}Y^{B}Y_{C}^{\dagger}
-6Y^{A}Y^{\dagger}_{B}Y^{B}Y_{A}^{\dagger}Y^{C}Y_{C}^{\dagger}
\big).
\end{align}
We choose real gamma matrices with the convention
$\gamma^2 = \gamma^0 \gamma^1$. An explicit representation would be
\begin{align}
\gamma^{0}=i\sigma^{2}, \quad \gamma^{1}=\sigma^{1}, \quad
\gamma^{2}=\sigma^{3}.
\end{align}

The action (\ref{ABJMa}) is known to be invariant under the 
${\cal N}=6$ supersymmetry
transformation~\cite{Aharony:2008ug,Hosomichi:2008qk,Gaiotto:2008cg,Terashima:2008sy},
\begin{align} \label{strf}
\delta Y^{A}&= i\omega^{AB}\psi_{B}, \nonumber\\
\delta \psi_{A}&= -\gamma_{\mu}\omega_{AB}D_{\mu}Y^{B}+\frac{2\pi}{k}\left[
-\omega_{AB}\big(Y^{C}Y_{C}^{\dagger}Y^{B}-Y^{B}Y_{C}^{\dagger}Y^{C}\big)
+2\omega_{BC}Y^{B}Y_{A}^{\dagger}Y^{C}\right] \nonumber \\
               &= -\gamma^{\mu}\omega_{AB}D_{\mu}Y^{B}
    +\omega_{BC} \left( \beta^{BC}_{\; A} 
                       + \delta_A^{[B} \beta^{C]D}_{\; D} \right), \nonumber\\
\delta A_\mu &= -\frac{2\pi}{k}\big(Y^{A}\psi^{B\dagger}\gamma_{\mu}
\omega_{AB}+\omega^{AB}\gamma_{\mu}\psi_{A}Y_{B}^{\dagger}\big), \nonumber \\
\delta \hat A_\mu &= \frac{2\pi}{k}\big(\psi^{A\dagger}Y^{B}\gamma_{\mu}
\omega_{AB}+\omega^{AB}\gamma_{\mu}Y_{A}^{\dagger}\psi_{B}\big),
\end{align}
where $\omega_{AB}$ are supersymmetry transformation parameters with
\begin{equation} \label{omegaab}
\omega^{AB} = (\omega_{AB})^* = -\frac12 \epsilon^{ABCD}\omega_{CD},
\end{equation}
and
\begin{align}
\beta^{AB}_{\; C}
  = \frac{4\pi}{k}Y^{[A}Y_{C}^{\dagger}Y^{B]}.
\label{bABC}
\end{align}

The form of the potential (\ref{potential0}) is manifestly SU$(4)$ invariant
but is not manifestly positive-definite. It can be written in a 
positive-definite form \cite{Bandres:2008ry,Hosomichi:2008ip} 
using the combination appearing in the second term of $\delta\psi_A$,
\begin{equation} \label{potential11}
V_0 = \frac23 \left|\beta^{BC}_{\;A} 
             + \delta^{[B}_A \beta^{C]D}_{\;D} \right|^2,
\end{equation}
where, for convenience, we have introduced the notation
$|{\cal O}|^2 \equiv {\rm tr} {\cal O}^\dagger {\cal O}$. 

There exists a unique mass deformation of the ABJM theory which respects 
the full ${\cal N}=6$ supersymmetry~\cite{Hosomichi:2008qk}. 
For the mass-deformed theory, the supersymmetric transformations (\ref{strf})
remain unchanged except the fermion fields for which there is an
additional transformation,
\begin{equation} \label{delmpsi}
\delta_{{\rm m}}\psi_{A}=\mu M_A^{\;B} \omega_{BC}Y^{C},
\end{equation}
where $\mu$ is the mass deformation parameter and 
$M_A^B = \rm{diag}(1,1,-1,-1)$. This reduces the R-symmetry from SU$(4)$ to
${\rm SU}(2) \times {\rm SU}(2) \times {\rm U}(1)$, and it leads to 
the following additional terms to the Lagrangian,
\begin{align}
\Delta V_{{\rm ferm}}&= {\rm tr}\, \mu \psi^{\dagger A}M_{A}^{\; B}\psi_{B},
\nonumber\\
\Delta V_0&={\rm tr}\left( \frac{4\pi\mu}{k}Y^{A}
Y^{\dagger}_{A}Y^{B}M_{B}^{\; C}Y_{C}^{\dagger}
-\frac{4\pi\mu}{k}Y^{\dagger}_{A}Y^{A}Y^{\dagger}_{B}M_{C}^{\; B}Y^{C}
+\mu^{2}Y^{\dagger}_{A}Y^{A} \right).
\label{md2}
\end{align} 
Combined with (\ref{potential11}), the potential $V_\textrm{m}$
in the mass-deformed 
theory can also be written in a manifestly positive-definite form,
\begin{equation} \label{potential3}
V_\textrm{m} = V_0+\Delta V_0 = \frac23 \left|\beta^{BC}_{\;A}
       + \delta^{[B}_A \beta^{C]D}_{\;D} + \mu M_A^{\;[B} Y_{}^{C]} \right|^2.
\end{equation}
This form is suitable for obtaining the half-BPS equations in the next section.

\subsection{Other formulations of mass-deformed theory}\label{ss21}
 
The ABJM theory can be described in terms of either 
the formalism of the component fields as above, ${\cal N}=1$ superfields
or ${\cal N}=2$ superfields. Depending on the formalism used, part of the
symmetry is hidden and the resulting forms of the potential look quite 
different from each other. 

This is also true for the mass-deformed theory.
The maximally supersymmetric mass-deformed theory given by
\eqref{delmpsi} and \eqref{md2} was first examined in terms of 
$\mathcal{N}=1$ superfield formalism \cite{Hosomichi:2008jb} and in 
component fields \cite{Gomis:2008vc}.  In addition to this mass deformation,
two other types of mass deformation have been proposed in 
${\cal N}=2$ superfield formalism in \cite{Gomis:2008vc}. 
They correspond to a D-term deformation and an F-term deformation respectively 
and seem to produce different theories having only $\mathcal N=2$ supersymmetry.
There is however a possibility that they have hidden symmetries not
manifest in $\mathcal N=2$ formalism and may actually result in the same
theory. Here we show that this is indeed the case. 
In other words, mass-deformed 
theories obtained by deforming D-term or F-term in $\mathcal N=2$ 
superfield formalism are the same as the one considered above with maximal
$\mathcal N=6$ supersymmetry.

Let us first start with ${\cal N}=1$ formalism.
Introducing the notation $Y^A = (Z^1, Z^2, W^{\dag1}, W^{\dag2})$,
$\mathcal{N}=1$ superpotential of ABJM theory is given as 
\begin{align}
\mathcal{W}_{\mathcal{N}=1}
&=
\frac{2\pi}{k}\mathrm{tr}\biggl(
\frac{1}{2}Z^{\dag}_{a}Z^{a}Z^{\dag}_{b}Z^{b}
-\frac{1}{2}Z^{a}Z^{\dag}_{a}Z^{b}Z^{\dag}_{b}
+\frac{1}{2}W_{a}W^{\dag a}W_{b}W^{\dag b}
-\frac{1}{2}W^{\dag a}W_{a}W^{\dag b}W_{b}
\notag\\
&\qquad\qquad{}
+Z^{a}Z^{\dag}_{a}W^{\dag b}W_{b}
-Z^{\dag}_{a}Z^{a}W_{b}W^{\dag b}
+2Z^{\dag}_{a}Z^{b}W_{b}W^{\dag a}
-2Z^{a}Z^{\dag}_{b}W^{\dag b}W_{a}
\biggr),
\label{n1sp}
\end{align}
where $a,b=1,2$. The bosonic potential can be written in the perfect 
square form 
\begin{equation}
V_0
=\mathrm{tr}\bigl(\hat{N}^{\dag}_{a}\hat{N}^{a}+\hat{M}^{\dag a}\hat{M}_{a}
\bigr) 
\label{n1bp}
\end{equation}
with
\begin{align}
\hat{N}^{a}
&=
-\frac{\partial\mathcal{W}_{\mathcal{N}=1}}{\partial Z^{\dag}_{a}}
\notag\\
&=
\frac{2\pi}{k}\bigl(Z^{b}Z^{\dag}_{b}Z^{a}-Z^{a}Z^{\dag}_{b}Z^{b}
-W^{\dag b}W_{b}Z^{a}+Z^{a}W_{b}W^{\dag b}
-2Z^{b}W_{b}W^{\dag a}+2W^{\dag a}W_{b}Z^{b}
\bigr),
\\
\hat{M}_{a}
&=
\frac{\partial\mathcal{W}_{\mathcal{N}=1}}{\partial W^{\dag a}}
\notag\\
&=
\frac{2\pi}{k}\bigl(W_{b}W^{\dag b}W_{a}-W_{a}W^{\dag b}W_{b}
+W_{a}Z^{b}Z^{\dag}_{b}-Z^{\dag}_{b}Z^{b}W_{a}
+2Z^{\dag}_{a}Z^{b}W_{b} - 2W_{b}Z^{b}Z^{\dag}_{a}
\bigr).
\end{align}
The SUSY-preserving mass deformation is introduced by 
the additional $\mathcal{N}=1$ superpotential as 
\begin{equation}
\Delta\mathcal{W}_{\mathcal{N}=1}=-\mu\, \mathrm{tr}\bigl(
Z^{\dag}_{a}Z^{a}-W^{\dag a}W_{a}
\bigr)
\end{equation}
which corresponds to the following replacement in the bosonic potential 
\eqref{n1bp}
\begin{equation}
\hat{N}^{a}\to \hat{N}^{a}+\mu Z^{a},\quad 
\hat{M}_{a}\to \hat{M}_{a}+\mu W_{a}. 
\end{equation}
Then the potential in the deformed theory is
\begin{equation} \label{potentialn1}
V_\textrm{m} = | \hat N^a + \mu Z^a |^2 + | \hat M_a + \mu W_a |^2
\end{equation}
which can be shown to be the same as \eqref{potential11}. 

In terms of $\mathcal{N}=2$ superfield formalism, the bosonic potential of 
ABJM theory is written as the sum of the D-term potential $V_{D}$ and 
the F-term potential $V_{F}$ \cite{Benna:2008zy}
\begin{equation}
V_0=V_{D}+V_{F}, 
\label{n2bp}
\end{equation}
where 
\begin{align}
V_{D}
&=
\mathrm{tr}\bigl(N^{\dag}_{a}N^{a}+M^{\dag a}M_{a}\bigr), 
\\
N^{a}
&=
\frac{2\pi}{k}\bigl(Z^{b}Z^{\dag}_{b}Z^{a}-Z^{a}Z^{\dag}_{b}Z^{b}
-W^{\dag b}W_{b}Z^{a}+Z^{a}W_{b}W^{\dag b}\bigr),
\\
M_{a}
&=
\frac{2\pi}{k}\bigl(W_{b}W^{\dag b}W_{a}-W_{a}W^{\dag b}W_{b}
+W_{a}Z^{b}Z^{\dag}_{b}-Z^{\dag}_{b}Z^{b}W_{a}\bigr),
\\[3mm]
V_{F}
&=\mathrm{tr}\bigl(F^{\dag}_{a}F^{a}+G^{\dag a}G_{a}\bigr), 
\\
F^{a}
&=\frac{4\pi}{k}\epsilon^{ac}\epsilon_{bd}W^{\dag b}Z^{\dag}_{c}W^{\dag d},
\\
G_{a}
&=-\frac{4\pi}{k}\epsilon_{ac}\epsilon^{bd}Z^{\dag}_{b}W^{\dag c}Z^{\dag}_{d}.
\end{align}
The F-term potential $V_{F}$ is obtained from 
the $\mathcal{N}=2$ superpotential $\mathcal{W}_{\mathcal{N}=2}$ as
\begin{gather}
\mathcal{W}_{\mathcal{N}=2}=\frac{2\pi}{k}\epsilon_{ac}\epsilon^{bd}
\mathrm{tr}\bigl(Z^{a}W_{b}Z^{c}W_{d}\bigr), 
\\[3mm]
F^{a}=\frac{\partial\mathcal{W}^{\dag}_{\mathcal{N}=2}}
{\partial Z^{\dag}_{a}},\quad 
G_{a}=\frac{\partial\mathcal{W}^{\dag}_{\mathcal{N}=2}}
{\partial W^{\dag a}}.
\end{gather}

In $\mathcal{N}=2$ superfield formalism, we can consider two kinds of mass 
deformations, D-term deformation and F-term deformation \cite{Gomis:2008vc}. 
The D-term deformation is introduced by a replacement in \eqref{n2bp} 
\begin{equation}
N^{a}\to N^{a}+\mu Z^{a},\quad M_{a}\to M_{a}+\mu W_{a}. 
\label{Dmd}
\end{equation}
Nevertheless one can explicitly verify that the resulting potential is 
the same as \eqref{potentialn1}, viz.,
\begin{equation} \label{equivalence}
\bigl|N^{a}+\mu Z^{a}\bigr|^{2}
+\bigl|M_{a}+\mu W_{a}\bigr|^{2}
+\bigl|F^{a}\bigr|^{2}+\bigl|G_{a}\bigr|^{2}
=\bigl|\hat{N}^{a}+\mu Z^{a}\bigr|^{2}
+\bigl|\hat{M}_{a}+\mu W_{a}\bigr|^{2} .
\end{equation}
Fermionic part can also be shown to be identical.
Hence the D-term deformation is the same as 
the maximally SUSY-preserving mass deformation. 
We note that the D-term deformation can be regarded as the 
Fayet-Illiopoulos term deformation 
when the gauge group is U($N$)$\times$U($N$) (not SU($N$)$\times$SU($N$)) 
\cite{Gomis:2008vc}. 

The other mass deformation is an F-term deformation which is introduced by 
the additional $\mathcal{N}=2$ superpotential 
\begin{equation}
\Delta\mathcal{W}_{\mathcal{N}=2}=\mu\,\mathrm{tr}\bigl(Z^{a}W_{a}\bigr).
\end{equation}
The deformation of bosonic potential is the form of \eqref{md2} 
with the off-diagonal mass matrix
\begin{equation}
M_{A}{}^{B}=
\begin{pmatrix}
0 & 0 & 1 & 0 \\
0 & 0 & 0 & 1 \\
1 & 0 & 0 & 0 \\
0 & 1 & 0 & 0 \\
\end{pmatrix}.
\label{offmass}
\end{equation}
By a field redefinition, this $M_{A}{}^{B}$ can be diagonalized and 
the F-term deformation is equivalent to the other deformations considered
above. In particular, $\mathcal{N}=6$ supersymmetry is still preserved 
in every case by deforming the transformation law as \eqref{delmpsi}.
At first sight, the F-term deformation looks different from the other 
deformations since they have ${\rm SU}(2)\times{\rm SU}(2)\times{\rm U}(1)$, 
while in the case of the F-term deformation only diagonal SU(2) can be 
seen. However we can find the extra SU(2) and U(1) symmetries in 
the F-term deformation. From the form of the mass matrix \eqref{offmass}, 
the generator of the extra SU(2) is obtained as
\begin{equation}
\begin{pmatrix}
0 & \frac{1}{2}\alpha_{i}\tau_{i} \\
\frac{1}{2}\alpha_{i}\tau_{i} & 0 
\end{pmatrix},
\end{equation}
and also the U(1) symmetry is generated by \eqref{offmass} itself. 
Here $\alpha_{i}\ (i=1,2,3)$ are the parameters of the extra SU(2) 
and $\tau_{i}$ are the Pauli matrices. Since both symmetries are generated by 
the off-diagonal matrices which mix $Z^{A}$ and $W^{\dagger A}$, 
their symmetries do not 
respect the structure of $\mathcal{N}=2$ superfield formalism 
(Recall that $Z^{A}$ is the lowest component of the chiral superfield 
whereas $W^{\dagger A}$ is the lowest component of 
the {\itshape anti}-chiral superfield). 
That is the reason why we can see only diagonal SU(2) 
in the F-term deformation. The fermionic mass term is also invariant 
under these extra SU(2) and U(1). 

Now we briefly discuss the vacua of the mass-deformed theory. From 
\eqref{potential3} the vacuum equation is given by
\begin{equation} \label{vacuumeq}
\beta^{BC}_{\;A} + \delta^{[B}_A \beta^{C]D}_{\;D} + \mu M_A^{\;[B} Y^{C]} = 0.
\end{equation}
Contracting with $\delta_C^A$ yields $ \beta^{BD}_{\;D} = \mu M_D^{\;B} Y^D $.
Inserting this into \eqref{vacuumeq}, we obtain
\begin{equation}
\beta^{BC}_{\;A} + \mu \left(
   \delta^{[B}_A M^{\;C]}_D Y^D + M_A^{\;[B} Y^{C]} \right) = 0.
\end{equation}
More explicitly, we have
\begin{align}
&\beta^{ab}_{\;a} + \mu Y^b = 0, \label{vacuumeq2}\\
&\beta^{pq}_{\;p} - \mu Y^q = 0, \label{vacuumeq3}\\
&\beta^{bp}_{\;a} =  \beta^{qa}_{\;p} = \beta^{pq}_{\;a} = \beta^{ab}_{\;p} =0,
\label{vacuumeq4}
\end{align}
where $a,b=1,2$ and $p,q=3,4$.
Equations \eqref{vacuumeq2}--\eqref{vacuumeq3} have been conjectured and 
analyzed in \cite{Gomis:2008vc}. Since \eqref{vacuumeq2} reduces to 
\eqref{vacuumeq3} with the substitution $Y^a \to Y^\dag_p$, we consider
only \eqref{vacuumeq3} to which there is essentially a unique irreducible 
solution,
\begin{equation} \label{irr}
Y^3_{mn} = \delta_{mn} \sqrt{\frac{k\mu}{2\pi}} \sqrt{m-1}\, ,\qquad
Y^4_{mn} = \delta_{m+1,n} \sqrt{\frac{k\mu}{2\pi}} \sqrt{N-n}\, .
\end{equation}
Then from \eqref{vacuumeq4} we see that $Y^1=Y^2=0$ identically
as claimed in \cite{Gomis:2008vc}.\footnote{We heard that
the same result was also obtained by~\cite{HLL}.}

\setcounter{equation}{0}
\section{Half-BPS Equations}\label{sec3}

Here we will obtain vortex-type half BPS equations in ABJM theory with and 
without mass deformation. First we consider the supersymmetric variation
of the fermions $\delta\psi^A = 0$ to obtain the BPS equations. 
Then we will get the energy bound by rewriting the energy functional 
in the form of complete squares. 

We impose the supersymmetric condition of the form
\begin{equation}
\gamma^{0}\omega_{AB}=is_{AB} \omega_{AB},\qquad s_{AB} = s_{BA} = \pm1,
\end{equation}
which reduces the number of supersymmetries by half.
Because of the property (\ref{omegaab}) among $\omega_{AB}$'s, we should have
$s_{34} = -s_{12}$, $s_{24} = -s_{13}$ and $s_{23} = -s_{14}$.
With the help of $\gamma^2 = -\gamma^1 \gamma^0$, the supersymmetric
variation of the fermion $\delta \psi_A$ can be reshuffled to
\begin{equation} 
\gamma^0 \delta \psi_A
  = \left[ \delta_A^{[B} D_0 Y^{C]}
          +\gamma^0 \left( \beta^{BC}_{\;A} + \delta^{[B}_A \beta^{C]D}_{\;D}
          + \mu M_A^{\;[B} Y^{C]} \right) \right] \omega_{BC} 
    -\gamma^2 \left(D_1 Y^B - \gamma^0 D_2 Y^B \right) \omega_{AB},
\end{equation}
where mass-deformed term (\ref{delmpsi}) has been included.
Then $\delta \psi_A = 0 $ implies that
\begin{align}\label{halfbps0}
&(D_1 - is_{AB} D_2)Y^B = 0, \nonumber \\
&\delta_A^{[B} D_0 Y^{C]}
          +is_{BC} \left( \beta^{BC}_{\;A} + \delta^{[B}_A \beta^{C]D}_{\;D}
          + \mu M_A^{\;[B} Y^{C]} \right) = 0,\qquad
\text{(no sum over $B,C$)}.
\end{align}

For nontrivial configurations at least one of $D_i Y^A$ should be nonzero.
Assume $D_1 Y^1 \neq 0$ for definiteness. Then from the first equation 
with $A=1$, it immediately follows that 
\begin{equation}
s_{12} = s_{13} = s_{14} \equiv s
\end{equation}
should be the same and 
\begin{equation} \label{halfbps1}
(D_1 - is D_2)Y^1 = 0.
\end{equation}
For $A\neq 1$, since $s_{23} = s_{34} = s_{42} = -s$, 
\begin{align}\label{halfbps2}
&(D_1 - is D_2)Y^A = 0 \qquad(A\neq 1), \nonumber \\
&(D_1 + is D_2)Y^A = 0 \qquad(A\neq 1),
\end{align}
where the first equation comes from $s_{A1} = s$ and the second from
$s_{AB} = -s$ for $A,B \neq 1$. Then
\begin{equation}\label{halfbps3}
D_iY^{A} = 0, \qquad (A\neq 1),
\end{equation}
and hence only one field can be nontrivial in half-BPS configurations.
This has also been obtained in \cite{Drukker:2008jm,SheikhJabbari:2009kr}.

Similarly, from the second line of (\ref{halfbps0}), 
we see that there are three different equations for each $D_0 Y^A$. They will
produce various constraints for consistency. Eventually
we end up with following equations:
\begin{alignat}{3}\label{halfbps4}
&D_0 Y^1 +is (\beta^{21}_{\;2} + \mu Y^1)  = 0, \qquad
&&D_0 Y^2 -is (\beta^{12}_{\;1} + \mu Y^2)  = 0, \nonumber \\
&D_0 Y^3 -is \beta^{13}_{\;1} = 0, \qquad
&&D_0 Y^4 -is \beta^{14}_{\;1} = 0, \nonumber \\
&\beta^{31}_{\;3} = \beta^{41}_{\;4} = \beta^{21}_{\;2} + \mu Y^1, \qquad
&&\beta^{43}_{\;4} = \mu Y^3, \qquad
\beta^{34}_{\;3} = \mu Y^4, \nonumber \\
&\beta^{32}_{\;3} = \beta^{42}_{\;4} = 
\beta^{23}_{\;2} = \beta^{24}_{\;2} = 0,&& \nonumber \\
&\beta^{BC}_{\;A} = 0 \qquad(A \neq B \neq C \neq A).&&
\end{alignat}
Equations (\ref{halfbps1}), (\ref{halfbps3}), (\ref{halfbps4}) form
the full set of half-BPS equations. In addition, Gauss' laws 
should also be satisfied,
\begin{align}
\frac{k}{2\pi}B=\frac{k}{2\pi} F_{12}=j^{0}
,\qquad
-\frac{k}{2\pi}{\hat B}=-\frac{k}{2\pi}{\hat F}_{12}={\hat j}^{0},
\label{Gl1}
\end{align}
where $j^0$ and $\hat j^0$ are respectively charge densities of the conserved
currents associated with U(1) rotations,
\begin{align} \label{jmu}
j_{\mu}=i(Y^A D_{\mu}Y_A^{\dagger}-D_{\mu}Y^A Y_A^{\dagger}),\nonumber \\
{\hat j}_{\mu}=i(Y_A^{\dagger}D_{\mu}Y^A - D_{\mu}Y_A^{\dagger}Y^A).
\end{align}

The half-BPS equations can also be obtained from the bosonic part of
the energy,
\begin{equation}
E = \int d^2x ( |D^0 Y_A|^2 + |D_i Y^A|^2 + V_\textrm{m} ),
\end{equation}
where the potential $V_\textrm{m}$ is given by (\ref{potential3}).
With the original form of the BPS equation (\ref{halfbps0}) in mind,
we can reshuffle the energy as
\begin{align} \label{energy1}
E &= \frac13 \int d^2x \left\{ 2\sum_{A,B,C} \left| \delta_A^{[B} D_0 Y^{C]}
          +is_{BC} \left( \beta^{BC}_{\;A} + \delta^{[B}_A \beta^{C]D}_{\;D}
          + \mu M_A^{\;[B} Y^{C]} \right) \right|^2 \right.
\nonumber\\
& \hspace{25mm}\left. 
     + \sum_{A\neq B}|(D_1 - is_{AB} D_2)Y^A|^2 \right\} 
\nonumber \\
  &\hspace{5mm} + is\, {\rm tr} \int d^2x \epsilon_{ij} \partial_i \left(
       Y_1^\dagger D_j Y^1 - \frac13 \sum_{A=2}^4 Y_A^\dagger D_j Y^A \right)
   - \frac{s}3\, \mu\, {\rm tr} 
                \int d^2x ( j^0 + 2 J_{12}^0 ),
\end{align}
where
\begin{equation}
J_{12}^0 = i(Y^1 D_0 Y_1^\dagger - D_0 Y^1 Y_1^\dagger)
         - i(Y^2 D_0 Y_2^\dagger - D_0 Y^2 Y_2^\dagger)
\end{equation}
is the charge density for an ${\rm SU}(4)$ rotation 
$Y^1 \rightarrow e^{-i\alpha}Y^1,\ Y^2 \rightarrow e^{i\alpha}Y^2$.
In obtaining this expression we have used the Gauss' laws \eqref{Gl1}.
Note that, for each and every $\mu$ and index $A$, 
$|D_\mu Y^A|^2$ is organized into three different complete 
squares in accordance with different supersymmetries and the factor 
$1/3$ in front of the integral accounts for the normalization. 

The first two absolute-square terms in (\ref{energy1}) precisely 
reproduce the half-BPS equations obtained before in \eqref{halfbps1}, 
\eqref{halfbps3}, and \eqref{halfbps4}. 
The first term in the last line is a boundary term%
\footnote{This bound can also be seen from the SUSY algebra obtained 
in~\cite{Low:2009kv}. }
which vanishes for well-behaved configurations. Then we get the energy bound
\begin{equation} \label{energy2}
E \ge  \frac13\, |\mu (Q + 2R_{12})|,
\end{equation}
where
$ Q = {\rm tr} \int d^2x\, j^0 $ and $ R_{12} = {\rm tr} \int d^2x\, J_{12}^0$.
The energy bound which is saturated for any well-behaved half-BPS 
configuration is proportional to the mass-deformation parameter $\mu$.
Note that in the energy bound there is the overall U(1) charge $Q$ 
in addition to the R-charge $R_{12}$ which also exists in BLG 
case~\cite{Hosomichi:2008qk}.

\setcounter{equation}{0}
\section{Solving Half-BPS Equations without Mass Deformation}\label{sec4}

Here we would like to solve half-BPS equations in the original ABJM theory 
$\mu=0$. In this case the equations are symmetric among $Y^2,Y^3,Y^4$. 

In the massless limit of $\mu \to 0$, the energy for half-BPS 
configurations is given by the total derivative term. From 
\eqref{energy1}, we have 
\begin{align}
E = \left|i\, {\rm tr} \int d^2x \epsilon_{ij} \partial_i \left(
       Y_1^\dagger D_j Y^1 - \frac13 \sum_{A=2}^4 Y_A^\dagger D_j Y^A 
\right)\right|
\label{ene11}
\end{align}
which vanishes for every well-behaved field configuration.
Therefore we expect that there would be no finite energy solution 
to the half-BPS equations other than vacuum configurations. Nevertheless
one can consider solutions with infinite energy, which may be physically
meaningful in the context of string theory.

The simplest solution would be obtained by assuming $Y^2=Y^3=Y^4=0$ for which
the only remaining equation is (\ref{halfbps1}). Magnetic fields vanish
as $D_0Y^A=0$. Then $Y^1$ is an arbitrary (anti-)holomorphic function
which can be singular as $Y^1 \sim z^{1/k}$ at the origin.
This solution has been discussed in the context of BLG theory with an M-theory 
interpretation \cite{Jeon:2008zj}. 

To obtain more nontrivial solutions at least one of $Y^2,Y^3,Y^4$ should be
nonzero. Due to the symmetry of the equations, we may assume $Y^2 \neq 0$.
Moreover with the help of the ${\rm U}(N) \times {\rm U}(N)$ gauge symmetry 
we can bring 
$Y^2$ to a diagonal form with increasing nonnegative real components,
\begin{equation} \label{y2}
Y^2 = \begin{pmatrix}
            v_1 I_{n_1} &             &        &             \\
                        & v_2 I_{n_2} &        &             \\
                        &             & \ddots &             \\
                        &             &        & v_k I_{n_k} 
      \end{pmatrix}, \qquad
(0 \le v_1 < v_2 < \cdots < v_k),
\end{equation}
where $\sum_{i=1}^k n_i = N$ and $I_{n_{i}}$ is the identity matrix of 
dimension $n_{i}$. 

We first concentrate on the constraint equations in (\ref{halfbps4}).
From the equation $\beta^{23}_{\;2} = \beta^{24}_{\;2} = 0$, 
it is not difficult to see that $Y^3$ and $Y^4$ are block diagonal,
\begin{equation}
Y^A = \begin{pmatrix}
            Y_{(1)}^A &           &        &             \\
                      & Y_{(2)}^A &        &             \\
                      &           & \ddots &             \\
                      &           &        & Y_{(k)}^A
      \end{pmatrix}, \qquad (A = 3,4).
\end{equation}
With these, 
$\beta^{24}_{\;3} = \beta^{43}_{\;2} = \beta^{32}_{\;3} = \beta^{42}_{\;4} = 0$
gives 
\begin{alignat}{2} \label{normal}
v_i [Y_{3(i)}^\dagger, Y_{(i)}^4 ] &= 0,\qquad &
v_i [Y_{(i)}^3, Y_{(i)}^4 ] &= 0, \nonumber \\
v_i [Y_{3(i)}^\dagger, Y_{(i)}^3 ] &= 0,\qquad &
v_i [Y_{4(i)}^\dagger, Y_{(i)}^4 ] &= 0,\qquad (\text{no sum over $i$}).
\end{alignat}
Note that all $v_i$'s are positive possibly except $v_1$ which we assume
nonzero for the moment. Then (\ref{normal}) implies that $Y^3$ and $Y^4$ are
normal matrices and commute to each other and their conjugates. Therefore, 
under a suitable unitary transformation, $Y^3$ and $Y^4$ become 
completely diagonal.
If $v_1 = 0$, we can utilize ${\rm U}(n_1) \times {\rm U}(n_1)$ symmetry 
to make $Y_{(1)}^3$ diagonal and reach the same conclusion.

Now we apply the remaining constraints, namely $\beta^{1B}_{\;A} = 0$
where $A,B=2,3,4$ and $A \neq B$. Since $Y^2,Y^3,Y^4$ are all diagonal,
this means $[Y^1, Y_A^\dagger Y^B] = 0$. If all $Y_A^\dagger Y^B$'s are
nondegenerate, $Y^1$ must be diagonal. But then all scalar fields are
diagonal and we will have only trivial solutions with vanishing magnetic field.
To obtain nontrivial solutions, there should be a common subspace where
{\em all} $Y_A^\dagger Y^B$'s are degenerate. Moreover, it is not difficult
to see that such a degenerate subspace should entirely belong to some
degenerate subspace of $Y^2$ in (\ref{y2}). This in turn means that
$Y^1$ can be at most block diagonal and in each block diagonal subspace
$Y^2, Y^3, Y^4$ are all proportional to the identity.

It is now sufficient to work within each subspace where $Y^A = v^AI$
$(A=2,3,4$). From (\ref{halfbps3}), we find that
$v^A$'s are constants and $A_i = \hat A_i$. $D_0 Y^A$'s are
determined from (\ref{halfbps4}),
\begin{align}
D_0 Y^1 &=0, \nonumber \\
D_0 Y^A &= -is \beta^{1A}_{\;1} = is \frac{2\pi}k v^A[Y^1,Y_1^\dagger],
\qquad  (A\neq 1).
\end{align}
Plugging this into the Gauss' laws \eqref{Gl1}, we are left with the 
following half-BPS equations without further constraint:
\begin{align} \label{reduced}
&(D_1 - is D_2)Y^1 = 0, \nonumber \\
&B = \hat B = -\frac{s}2 \left( \frac{2\pi v}{k} \right)^2 [Y^1,Y_1^\dagger],
\end{align}
where $v^2 = \sum_{A=2}^4|v^A|^2$ is a positive constant.
Note that this result is completely general without any ansatz employed.

The equation (\ref{reduced}) is not entirely new. It can be obtained as the 
half-BPS equation of the super Yang-Mills theory with coupling constant
\begin{equation}
g_\textrm{YM}=\frac{2\pi v}{k}. \label{coupling}
\end{equation}
This identification has already appeared in the context of the
compactification of BLG/ABJM theory (from M2 to D2) 
\cite{Mukhi:2008ux,Pang:2008hw}. 
For finite $k$, there are correction terms to Yang-Mills Lagrangian.
However here we do not need to take the limit $v,\,k\to\infty$ as long
as the half-BPS equation is concerned. 

Alternatively, if the sign of the second equation in \eqref{reduced}
is flipped, it is exactly the same as the half-BPS equation in
{\em nonrelativistic} Chern-Simons theory with an adjoint
matter~\cite{Grossman:1990it,Dunne:1990qe} where the solutions have
been studied extensively. Here we briefly describe some simple solutions
of \eqref{reduced} with $s=1$ for definiteness. 
Introducing complex notations $z=x_{1}+ix_{2}$ and $A=(A_{1}-iA_{2})/2$,
we take the ansatz,
\begin{align}\label{ansz1}
Y^1&=\sum_{a=1}^{N-1} y_a e^a + y_M E^{-M}, 
\nonumber \\
A &= \sum_{a=1}^{N-1} A_a h^a, \qquad 
\bar A = \sum_{a=1}^{N-1} A^*_a h^a,
\end{align}
where $h^a$ and $e^a$ are SU($N$) generators in the Chevalley basis satisfying
$[e^a,\, e^{-a'}] = \delta_{a a'} h^a$,  $[h^a,\, e^b]= K_{a b} e^b$
with $K_{a b}$ being the Cartan matrix, and 
$E^{-M}$ is the Hermitian conjugate of the maximal ladder operator $E^M$
commuting with $e^a$'s. 
Plugging (\ref{ansz1}) into (\ref{reduced}), we obtain (affine-)Toda-type 
equation,
\begin{align}\label{udeqn}
\partial \bar\partial \ln |y_a|^2 &= 4v \left(\frac{2\pi}{k}\right)^2
\sum_{b=1}^{N-1} K_{a b}\left( |y_b|^2 - 
\frac{|G(z)|^2}{|c_b|^2\prod_{c=1}^{N-1} |y_c|^2}\right), 
\nonumber \\
y_M &= \frac{G(z)}{\prod_{a=1}^{N-1} y_a}, 
\end{align}
where $G(z)$ is an arbitrary holomorphic function. For SU(2), this reduces to
Liouville-type equation (with $G=0$) or Sinh-Gordon-type equation 
(with $G=$const.) considered in \cite{Jeon:2008zj}
in the context of BLG theory. The solutions however all have to have infinite
energy as we mentioned before. This is also consistent with the fact that
\eqref{reduced} is obtained from super Yang-Mills theory without symmetry 
breaking potential.

\setcounter{equation}{0}
\section{Solving Half-BPS Equations in the Mass-Deformed Theory}\label{sec5}

In this section we solve half-BPS equations in \eqref{halfbps1}, 
\eqref{halfbps3}, \eqref{halfbps4} in the mass-deformed theory. 
In this case, the constraint equations in (\ref{halfbps4}) are more 
complicated and we first consider ${\rm U}(2) \times {\rm U}(2)$ case,
and then discuss general U($N$)$\times$U($N$) case.

\subsection{${\rm U}(2) \times {\rm U}(2)$}

From the constraint $\beta^{43}_{\;4} = \mu Y^3$ and
$\beta^{34}_{\;3} = \mu Y^4$, it is clear that $Y^3$ is nonzero if and only
if $Y^4$ is nonzero. Let us first consider the case that both are nonzero.
Utilizing the gauge symmetry, we may assume without loss of generality that
$Y^3$ is diagonal with nonnegative real entries,
\begin{equation}
Y^3 = \sqrt{ \frac{k |\mu|}{2\pi} } 
      \begin{pmatrix} c & 0 \\ 0 & d \end{pmatrix},\qquad 0 \le c \le d.
\end{equation}
Suppose $c\neq d$. Then solving the two constraints as well as 
$\beta^{BC}_{\;A}=0$, we find only a trivial solution that $Y^1 = Y^2 = 0$ 
and\footnote{Hereafter we assume that $\mu>0$. Negative $\mu$ case can be 
analyzed in a similar manner and the same results are obtained except that the 
first and the second components are exchanged.}
\begin{equation}
Y^3 = \sqrt{ \frac{k \mu}{2\pi} } 
      \begin{pmatrix} 0 & 0 \\ 0 & d \end{pmatrix},\qquad
Y^4 = \sqrt{ \frac{k \mu}{2\pi} } 
      \begin{pmatrix} 0 & 0 \\ e^{i\chi} & 0 \end{pmatrix},
\end{equation}
where $\chi$ is a real constant.

To obtain nontrivial solutions $Y^3$ must be proportional to the identity. 
But then from the constraint $\mu Y^4 = \beta^{34}_{\;3}$, we see that 
$Y^4=0$ which in turn implies $Y^3=0$. Therefore we are left with only 
one constraint,
\begin{equation} \label{constraint9}
\beta^{21}_{\;2} + \mu Y^1 = 0.
\end{equation}
Note that this constraint with the first BPS equation in \eqref{halfbps4}
implies $D_0Y^1=0$ and hence only $Y^2$ is responsible for
the charge of which the U(1) current is given in \eqref{jmu}, 
while $Y^1$ satisfying \eqref{halfbps1} gives the vorticity. 
Nonzero charge due to $Y^2$ is
then related to the magnetic field through the Gauss laws \eqref{Gl1} 
which are a characteristic nature of the Chern-Simons gauge theory.

As before, we can assume that $Y^2$ is a diagonal matrix with nonnegative
increasing real entries. Solving (\ref{constraint9}) gives
\begin{equation} \label{n2y1y2}
Y^1 = \sqrt{ \frac{k \mu}{2\pi} } 
      \begin{pmatrix} 0 & f \\ 0 & 0 \end{pmatrix},\qquad
Y^2 = \sqrt{ \frac{k \mu}{2\pi} } 
      \begin{pmatrix} a & 0 \\ 0 & \sqrt{a^2+1} \end{pmatrix},\qquad (a\ge0).
\end{equation}
Inserting this into the equation for $D_0 Y^2$ in \eqref{halfbps4} and 
Gauss's laws \eqref{Gl1}, we find that the magnetic fields are given by
\begin{equation}
B = \hat B = -2s \mu^2 \begin{pmatrix}
  a^2(1+|f|^2) & 0 \\ 0 & (a^2+1)(1-|f|^2) \end{pmatrix},
\label{BBh}
\end{equation}
which means that the gauge fields are diagonal,
\begin{equation}
A_i = \begin{pmatrix}  u_i & 0 \\ 0  & v_i \end{pmatrix}.
\end{equation}
Note that $\hat A_i$ is the same as $A_i$ up to a gauge degree which 
can be put to zero.
Then from $D_i Y^2=0$, $a$ in \eqref{n2y1y2} must be a constant. 

Finally, from the equation $(D_1-isD_2)Y^1=0$ in \eqref{halfbps1}, 
we can express the
gauge fields in terms of $Y^1$. Explicitly, with $s=-1$ for definiteness,
\begin{align} \label{uvf}
\bar u - \bar v = i \bar\partial \ln f.
\end{align}
Comparing this with the magnetic field
$B = \frac2i(\partial \bar A - \bar\partial A)$, we obtain
\begin{align} \label{mh}
\partial \bar\partial \ln |f|^2 
+i(\partial{\bar \partial}-{\bar \partial}\partial )\Omega
=  \mu^2 \left[ (2a^2+1)|f|^2 -1 \right],
\end{align}
where $\Omega$ is the phase of the scalar field, $f=|f|e^{i\Omega}$.
This is the celebrated vortex equation appearing in Maxwell-Higgs theory and
has been extensively studied~\cite{Taubes:1979tm}. The same equation has
also obtained in mass-deformed BLG theory~\cite{Kim:2008cp}. Note however
that, although the final equation is the same, U(1) part plays a nontrivial
role in the present case.

Let us now calculate the energy of the solution satisfying (\ref{mh}).
Since $D_0 Y^1 = 0$, $J^0_{12} = j^0 = \frac{k}{2\pi}B$ and hence
\begin{equation}
E = \left| \frac{k \mu}{2\pi} {\rm tr}\int d^2x\,B \right|.
\end{equation}
Therefore the energy is given by the trace part of the magnetic field
while the vorticity of the solution comes from the relative part as seen in
\eqref{uvf}. Then the energy of a vortex solution is not proportional 
to the vorticity in general and can be infinite for some solutions.
Using \eqref{BBh}, we can rewrite
\begin{equation}
E = \left| \frac{k \mu}{2\pi} \int d^2x\, 2\mu^2 (2a^2 + 1 - |f|^2)  \right|.
\end{equation}
This implies that, for finite energy solutions, 
$f$ should behave asymptotically as
\begin{equation}
|f|^2 \longrightarrow 2a^2 + 1,\qquad r\longrightarrow \infty,
\end{equation}
which is consistent with (\ref{mh}) only for $a=0$.
In other words, to obtain finite energy solutions we should set $a=0$.
In this case, $u_i = 0$ and using (\ref{uvf}), we find that the magnetic
flux is given by
\begin{equation}
{\rm tr}\int d^2x\, B = 2\pi n, \qquad(n \in \mathbb{Z}),
\end{equation}
where $n$ is the vorticity.
Then the energy becomes
\begin{equation} \label{u2energy}
E = nk\mu .
\end{equation}

A characteristic nature of configurations in Chern-Simons gauge theory is that 
they usually carry nonzero angular momentum. However in this case it vanishes.
To see this note that the linear momentum density is proportional to the
combination $ D_0Y^A D_iY_A^\dag $. However for the present case either
$D_0Y^A$ vanishes or $D_iY^A$ vanishes because fields do not carry both
charge and vorticity as discussed above. 

When $a \neq 0$, the solution may be interpreted as vortices in the presence
of a constant magnetic field with the energy written as a sum
of the vortex part and the constant part,
\begin{equation}
E = \frac{nk\mu}{2a^2 +1}
        + \int d^2x\, \frac{4k \mu^3 a^2 (a^2 +1)}{\pi(2a^2+1)}.
\end{equation}

It would be illuminating to examine the origin of the 
Maxwell-Higgs vortex equation  
in the Chern-Simons gauge theory which has a sextic potential having 
a minimum at the origin. For this purpose we consider the ansatz
\begin{equation} \label{fgansatz}
Y^1 = \sqrt{ \frac{k \mu}{2\pi} } 
      \begin{pmatrix} 0 & f \\ 0 & 0 \end{pmatrix},\qquad
Y^2 = \sqrt{ \frac{k \mu}{2\pi} } 
      \begin{pmatrix} 0 & 0 \\ 0 & g \end{pmatrix},\qquad
Y^3=Y^4=0,
\end{equation}
and calculate the form of the potential as a function of $f$ and $g$.
With $g=1$ this reduces to \eqref{n2y1y2} so we would like to see
how the potential changes as $g$ changes. Plugging \eqref{fgansatz} into
\eqref{potential3}, we have the following potential in the mass-deformed 
theory,
\begin{equation} \label{rvm}
V_\textrm{m} (f,g) = \left(\frac{k}{2\pi}\right)^2 \mu^3 
                     [ |f|^2 ( |g|^2-1 )^2 + |g|^2 ( |f|^2-1 )^2 ].
\end{equation}
Then $V_\textrm{m}$ vanishes at $f=g=0$ and $|f|=|g|=1$ as it should be.
{}From this potential we get the quartic potential 
$(|f|^2-1)^2$ with $g=1$ which is the potential appearing in Maxwell-Higgs
theory. Note that $f=0,\ g=1$ is {\em not} a local maximum of the potential
since $V_\textrm{m}(f=0,g) \sim |g|^2$. One may wonder how the
configuration does not roll down to the origin. This is due to the
special nature of the Gauss law in Chern-Simons gauge theory, 
namely the magnetic
field is proportional to the charge density \eqref{Gl1}. Replacing $D_0Y^2$
by the magnetic field in the energy expression, we obtain an effective
potential term $|B/g|^2$ which acts as a barrier at the origin 
$(g\rightarrow 0)$. 
This can be interpreted as a centrifugal term inversely proportional to 
$1/g^{2}$ due to the rotation in $Y^2$ 
plane.

Along the direction $f=g$, \eqref{rvm} becomes the sextic potential 
$|f|^2(|f|^2-1)^2$ which appears in U(1) self-dual Chern-Simons matter 
system~\cite{Hong:1990yh}. It turns out that this direction corresponds
to a less supersymmetric BPS case and will be reported elsewhere~\cite{GKKKN}.

\subsection{{\rm U}($N$)$\times${\rm U}($N$)}
As in the previous subsection we start with the constraint
$\beta^{43}_{\;4} = \mu Y^3$ and $\beta^{34}_{\;3} = \mu Y^4$.
Solutions of these equations have already been considered in 
the end of section~\ref{sec2}.
For the irreducible solution \eqref{irr},
it is easy to show that the other constraints produce only a trivial
solution $Y^1 = Y^2 = 0$. Therefore to obtain nontrivial solutions we should
consider reducible cases. Since there are many different possibilities,
here we will analyze only some representative cases.

First note that the vorticity is carried by the field $Y^1$ and the other
scalars constrain the degrees of freedom of $Y^1$ through constraint 
equations in (\ref{halfbps4}). Then in order to find nontrivial solutions
it is desirable to assume that $Y^2,Y^3,Y^4$ take simple forms. However,
we cannot put all of these to be zero because the magnetic field would vanish
in that case. Furthermore, if one of $Y^3,Y^4$ vanishes, the other should also
vanish from the last two constraints in the third line of (\ref{halfbps4}).
Considering all these, it is natural to set $Y^3=Y^4=0$ while $Y^2\neq 0$. 
Then we are left with only one constraint $\beta^{21}_{\;2} + \mu Y^1 = 0$
as in the previous subsection.

An interesting nontrivial solution is obtained with the ansatz which 
generalizes (\ref{n2y1y2}) with $a=0$ to a block matrix form,
\begin{equation} \label{ny1y2}
Y^1 = \sqrt{\frac{k\mu}{2\pi}}
      \begin{pmatrix} 0_{N_1\times N_1} & F_{N_1\times N_2} \\
                      0_{N_2\times N_1} & 0_{N_2\times N_2} \end{pmatrix},\qquad
Y^2 = \sqrt{\frac{k\mu}{2\pi}}
\begin{pmatrix} 0_{N_1\times N_1} & 0_{N_1\times N_2} \\
                      0_{N_2\times N_1} & G_{N_2\times N_2} \end{pmatrix},
\end{equation}
where $N_1, N_2$ are positive integers satisfying $N_1 + N_2=N$
and the subscript denotes the dimensionality of each block 
which we omit from now on. We also only consider the case $N_1 \ge N_2$
for simplicity.
With the help of gauge symmetry, $G$ can be chosen to be a diagonal matrix
with real nonnegative entries. Solving the constraint,
we find $F=FG^\dagger G$ which implies $G$ is the $N_2$-dimensional 
identity matrix. Then the magnetic field is given by
\begin{equation} \label{bn2}
B = \hat B = -2s \mu^2
             \begin{pmatrix} 0 & 0 \\ 0 & I - F^\dagger F \end{pmatrix},
\end{equation}
and up to a gauge we can write
\begin{equation}
A_i = \hat A_i = \begin{pmatrix} 0 & 0 \\ 0 & \tilde A_i \end{pmatrix}.
\end{equation}
The only remaining equation is (\ref{halfbps1}) and it becomes
\begin{equation}
(D_1 + is D_2) F^\dagger = 0,
\end{equation}
where $D_i F^\dagger = \partial _i F^\dagger + i \tilde A_i F^\dagger$.
Together with (\ref{bn2}) this forms the nonabelian vortex equation in
U($N_2$) Yang-Mills theory with $N_1$ fundamental scalar 
fields and has been studied 
extensively~\cite{Hanany:2003hp, Eto:2006pg}. If $N=2$
and $N_1=N_2=1$, it reduces to the abelian vortex equation
obtained in the previous subsection.
For this configuration,
we have $j^{0}=J^{0}_{12} = \frac{k}{2\pi}B$ and obtain the energy as
\begin{align}
E = \left| \frac{k \mu}{2\pi} {\rm tr}\int d^2x\,B \right| = nk\mu,
\end{align}
which is the generalization of \eqref{u2energy}.

The ansatz (\ref{ny1y2}) with $G=I$ may be considered as a solution to
$\beta^{21}_{\;2} + \mu Y^1 = 0$ with maximal degeneracy. 
The irreducible nondegenerate ansatz similar to (\ref{irr}) is
\begin{equation}
Y^1_{mn} = \delta_{m+1,n} \sqrt{\frac{k\mu}{2\pi}} f_m, \qquad
Y^2_{mn} = \delta_{mn} \sqrt{\frac{k\mu}{2\pi}} a_m,
\end{equation}
where 
\begin{equation}
a_m = \sqrt{a_1^2 + m-1}\, .
\end{equation}
Here $a_1$ is a nonnegative constant and 
$f_1,\ldots,f_{N-1}$ are functions to be determined. 
The irreducible vacuum would be obtained for $a_1=0$ and $f_n=\sqrt{N-n}$.
With this ansatz the magnetic field becomes a diagonal matrix given by
\begin{equation}
B_{mn} = \hat B_{mn} = -2s\mu^2 a_m^2 ( |f_m|^2 - |f_{m-1}|^2 + 1 ) \delta_{mn},
\end{equation}
where $f_0 = f_N = 0$. Eliminating the gauge fields from (\ref{halfbps1}),
we obtain $N-1$ coupled differential equations,
\begin{equation} \label{nbps}
\partial\bar\partial \ln|f_m|^2 = -\mu^2[
  a_m^2 |f_{m-1}|^2 - ( a_m^2 + a_{m+1}^2) |f_m|^2 + a_{m+1}^2 |f_{m+1}|^2 +1 ].
\end{equation}
This type of coupled equations has appeared in U$(1)^{N-1}$ gauge theories with 
$N-1$ Higgs fields which couple to the gauge fields~\cite{Kim:1993mh}.

As in U(2)$\times$U(2) case, the solutions of \eqref{nbps} do not necessarily
have finite energy. 
For the finite energy in (\ref{energy2}), the trace of the magnetic
field should vanish in the asymptotic limit $r \rightarrow \infty$.
It is not difficult to find that the condition is consistent 
with the asymptotic behavior obtained from (\ref{nbps}) only when
the constant $a_1$ vanishes. Otherwise we would have
infinite energy configurations with background of a constant 
magnetic field as discussed in the previous subsection.

From now on we consider only the case $a_1=0$ for which \eqref{nbps}
reduces to
\begin{equation} \label{nbps2}
\partial\bar\partial \ln|f_m|^2 = -\mu^2[
  (m-1) |f_{m-1}|^2 - (2m-1) |f_m|^2 + m |f_{m+1}|^2 +1 ].
\end{equation}
The asymptotic value of $|f_m|$'s are determined by requiring that
the right hand side of \eqref{nbps2} vanish,
\begin{equation} \label{afn}
|f_m| \rightarrow \sqrt{N-m}, \qquad r \rightarrow \infty.
\end{equation}
Note that this is nothing but the irreducible vacuum values as it should be.
An obvious solution of \eqref{nbps2} is obtained with the ansatz
\begin{equation} \label{fnf}
f_m = \sqrt{N-m}f,
\end{equation}
which is consistent with \eqref{afn}. Then \eqref{nbps2} reduces to a
single equation,
\begin{equation}
\partial\bar\partial \ln|f|^2 = -\mu^2 ( 1 - |f|^2 ),
\end{equation}
which is again the Maxwell-Higgs vortex equation which generalizes the result
of U(2)$\times$U(2) case in the previous subsection.

The ansatz \eqref{fnf} assumes that all the components $f_m$'s have the same
functional form and, in particular, the same vorticity. There are
however more general solutions for which $f_m$'s carry different 
vorticities~\cite{Kim:1993mh}. Let $n_m$ be the vorticity of $f_m$. Then
the energy can be calculated by taking the trace of the magnetic field,
\begin{equation}
E = \left| \frac{k \mu}{2\pi} {\rm tr}\int d^2x\,B \right|
  = k\mu \left|\sum_{l=2}^N \sum_{m=1}^{l-1} n_m \right|.
\end{equation}
If $n_m=n$ are the same for all $m$, the energy reduces to
$ E = nk\mu N(N-1)/2$.

Other than the cases considered above, we have tried some other ansatz on which
we briefly comment here. For U(3)$\times$U(3), we worked out the
equations in the most general way including reducible cases. In most
cases the result is essentially some embedding of U(2)$\times$U(2) case.
When $Y^2=0$ while $Y^3,Y^4$ are not zero, Liouville-type equation 
can also be obtained as in the previous section. For $N>3$,
we did not fully analyzed the constraints but it seems that most cases
fall into one of those considered here.

\setcounter{equation}{0}
\section{Summary and Outlook}\label{sec6}

In this paper, we investigated vortex-type half-BPS equations in the ABJM
theory with or without mass deformation 
We obtained the energy bound (\ref{energy2})
which is proportional to the mass-deformation parameter. 
We also showed that the D-term deformation and the F-term deformation are 
the same as the mass deformation preserving maximal 
$\mathcal{N}=6$ supersymmetry.

For the undeformed ABJM theory, we solved all the constraint equations 
in the BPS equations. The resulting equation is shown to be the 
half-BPS equation in supersymmetric Yang-Mills theory.
It has no finite energy regular solution.

In the mass-deformed theory, we showed that the BPS equations for 
U(2)$\times$U(2) case reduce to the vortex equation appearing in Maxwell-Higgs
theory which is known to have multi-vortex solutions. We obtained pure
vortex solutions with the energy given by its vorticity as well as
vortices in the constant background of magnetic field. We explored the origin
of Maxwell-Higgs vortex in the Chern-Simons gauge theory. For U($N)\times$U($N$)
case with $N>2$, we obtained the nonabelian vortex equation of
Yang-Mills-Higgs theory and also more general equations. 
It would be interesting to study the moduli space of these solutions.

There are many issues not addressed in this paper. 
It is straightforward to extend our analysis to the cases with less 
supersymmetry.
A notable case among them is the $\mathcal{N}=1$ BPS equation which
turns out to reduce to the vortex equation in U(1) Chern-Simons-Higgs 
system~\cite{Hong:1990yh}. This has been also considered in \cite{Arai:2008kv}
in the context of F-term deformation.

Since the ABJM theory is defined on a $\mathbb{Z}_k$ orbifold with
the action $Y^A \rightarrow e^{2\pi i/k} Y^A$, one may explore the possibility
of configurations having fractional vorticity with phase dependence
of the form $Y^A \sim e^{i\theta/k}$. It can be shown that this is possible
in less supersymmetric solutions such as $\mathcal{N}=1$ case~\cite{GKKKN}.

In this paper, we considered the theory purely from the viewpoint of a field 
theory and did not attempt to interpret the solutions in the context 
of M-theory. We would like to investigate these issues in the forthcoming
publication~\cite{GKKKN}.

\section*{Acknowledgments}
The authors thank Ki-Myeong Lee and Sungjay Lee for useful discussions. 
H.~N. thanks Masato Arai, Giacomo Marmorini and Muneto Nitta for 
helpful discussions. 
This work was supported by Astrophysical Research
Center for the Structure and Evolution of the Cosmos (ARCSEC)) (O.K.)
and grant No. R01-2006-000-10965-0 from the Basic Research Program
through the Korea Science $\&$ Engineering Foundation (Y.K.,H.N.).
This work was also supported by the Science Research Center Program of the
Korea Science and Engineering Foundation through the Center for
Quantum Spacetime (CQUeST) with grant number R11-2005-021 
and the World Class University grant R32-2008-000-10130-0 (C.K.).


\begin{thebibliography}{100}

\bibitem{Townsend:1996xj}
For a review, see P.~K.~Townsend,
``Four lectures on M-theory,''
arXiv:hep-th/9612121.

\bibitem{Bagger:2006sk}
J.~Bagger and N.~Lambert,
``Modeling multiple M2's,''
Phys.\ Rev.\  D {\bf 75}, 045020 (2007)
[arXiv:hep-th/0611108];
``Gauge Symmetry and Supersymmetry of Multiple M2-Branes,''
Phys.\ Rev.\  D {\bf 77}, 065008 (2008)
[arXiv:0711.0955 [hep-th]];
``Comments On Multiple M2-branes,''
JHEP {\bf 0802}, 105 (2008)
[arXiv:0712.3738 [hep-th]].

\bibitem{Gustavsson:2007vu}
  A.~Gustavsson,
  ``Algebraic structures on parallel M2-branes,''
  Nucl.\ Phys.\  B {\bf 811}, 66 (2009)
  [arXiv:0709.1260 [hep-th]].

\bibitem{Aharony:2008ug}
  O.~Aharony, O.~Bergman, D.~L.~Jafferis and J.~Maldacena,
  ``N=6 superconformal Chern-Simons-matter theories, M2-branes and their
  gravity duals,''
  JHEP {\bf 0810}, 091 (2008)
  [arXiv:0806.1218 [hep-th]].

\bibitem{Basu:2004ed}
A.~Basu and J.~A.~Harvey,
``The M2-M5 brane system and a generalized Nahm's equation,''
Nucl.\ Phys.\  B {\bf 713}, 136 (2005)
[arXiv:hep-th/0412310].

\bibitem{Hosomichi:2008qk}
  K.~Hosomichi, K.~M.~Lee and S.~Lee,
  ``Mass-Deformed Bagger-Lambert Theory and its BPS Objects,''
  Phys.\ Rev.\  D {\bf 78}, 066015 (2008)
  [arXiv:0804.2519 [hep-th]].

\bibitem{Krishnan:2008zm}
C.~Krishnan and C.~Maccaferri,
``Membranes on Calibrations,''
JHEP {\bf 0807}, 005 (2008)
[arXiv:0805.3125 [hep-th]].

\bibitem{Bonelli:2008kh}
  G.~Bonelli, A.~Tanzini and M.~Zabzine,
  Phys.\ Lett.\  B {\bf 672} (2009) 390
  [arXiv:0807.5113 [hep-th]].

\bibitem{Terashima:2008sy}
  S.~Terashima,
  ``On M5-branes in N=6 Membrane Action,''
  JHEP {\bf 0808}, 080 (2008)
  [arXiv:0807.0197 [hep-th]];

  K.~Hanaki and H.~Lin,
  ``M2-M5 Systems in N=6 Chern-Simons Theory,''
  JHEP {\bf 0809}, 067 (2008)
  [arXiv:0807.2074 [hep-th]].

\bibitem{Hong:1990yh}
  J.~Hong, Y.~Kim and P.~Y.~Pac,
  ``On The Multivortex Solutions Of The Abelian Chern-Simons-Higgs Theory,''
  Phys.\ Rev.\ Lett.\  {\bf 64}, 2230 (1990);
  
  R.~Jackiw and E.~J.~Weinberg,
  ``Selfdual Chern-Simons Vortices,''
  Phys.\ Rev.\ Lett.\  {\bf 64}, 2234 (1990).

\bibitem{Lee:1990it}
C.~Lee, K.~M.~Lee and E.~J.~Weinberg,
``Supersymmetry And Selfdual Chern-Simons Systems,''
Phys.\ Lett.\  B {\bf 243}, 105 (1990).

\bibitem{Jackiw:1990pr}
R.~Jackiw, K.~M.~Lee and E.~J.~Weinberg,
``Selfdual Chern-Simons solitons,''
Phys.\ Rev.\  D {\bf 42}, 3488 (1990).

\bibitem{Kim:1992uw}
C.~Kim,
``Selfdual vortices in the generalized Abelian Higgs model with independent
Chern-Simons interaction,''
Phys.\ Rev.\  D {\bf 47}, 673 (1993)
[arXiv:hep-th/9209110].

\bibitem{Cugliandolo:1990nb}
L.~F.~Cugliandolo, G.~Lozano, M.~V.~Manias and F.~A.~Schaposnik,
``Bogomolny equations for nonAbelian Chern-Simons Higgs theories,''
Mod.\ Phys.\ Lett.\  A {\bf 6}, 479 (1991);
%
K.~M.~Lee,
``Selfdual nonabelian Chern-Simons solitons,''
Phys.\ Rev.\ Lett.\  {\bf 66}, 553 (1991);
%
``Relativistic nonAbelian selfdual Chern-Simons systems,''
Phys.\ Lett.\  B {\bf 255}, 381 (1991).

\bibitem{Gomis:2008vc}
J.~Gomis, D.~Rodriguez-Gomez, M.~Van Raamsdonk and H.~Verlinde,
``A Massive Study of M2-brane Proposals,''
JHEP {\bf 0809}, 113 (2008)
[arXiv:0807.1074 [hep-th]].

\bibitem{Jeon:2008bx}
  I.~Jeon, J.~Kim, N.~Kim, S.~W.~Kim and J.~H.~Park,
  ``Classification of the BPS states in Bagger-Lambert Theory,''
  JHEP {\bf 0807}, 056 (2008)
  [arXiv:0805.3236 [hep-th]].

\bibitem{Jeon:2008zj}
  I.~Jeon, J.~Kim, B.~H.~Lee, J.~H.~Park and N.~Kim,
  ``M-brane bound states and the supersymmetry of BPS solutions in the
  Bagger-Lambert theory,''
  arXiv:0809.0856 [hep-th].

\bibitem{Kim:2008cp}
J.~Kim and B.~H.~Lee,
``Abelian Vortex in Bagger-Lambert-Gustavsson Theory,''
JHEP {\bf 0901}, 001 (2009)
[arXiv:0810.3091 [hep-th]].

\bibitem{Arai:2008kv}
  M.~Arai, C.~Montonen and S.~Sasaki,
  ``Vortices, Q-balls and Domain Walls on Dielectric M2-branes,''
  JHEP {\bf 0903}, 119 (2009)
  [arXiv:0812.4437 [hep-th]].

\bibitem{Drukker:2008jm}
  N.~Drukker, J.~Gomis and D.~Young,
  ``Vortex Loop Operators, M2-branes and Holography,''
  JHEP {\bf 0903}, 004 (2009)
  [arXiv:0810.4344 [hep-th]].

\bibitem{Mukhi:2008ux}
S.~Mukhi and C.~Papageorgakis,
``M2 to D2,''
JHEP {\bf 0805}, 085 (2008)
[arXiv:0803.3218 [hep-th]].

\bibitem{Pang:2008hw}
  Y.~Pang and T.~Wang,
  ``From N M2's to N D2's,''
  Phys.\ Rev.\  D {\bf 78}, 125007 (2008)
  [arXiv:0807.1444 [hep-th]].

\bibitem{Gaiotto:2008cg}
  D.~Gaiotto, S.~Giombi and X.~Yin,
  ``Spin Chains in N=6 Superconformal Chern-Simons-Matter Theory,''
  JHEP {\bf 0904}, 066 (2009)
  [arXiv:0806.4589 [hep-th]].

\bibitem{Bandres:2008ry}
  M.~A.~Bandres, A.~E.~Lipstein and J.~H.~Schwarz,
  ``Studies of the ABJM Theory in a Formulation with Manifest SU(4)
  R-Symmetry,''
  JHEP {\bf 0809}, 027 (2008)
  [arXiv:0807.0880 [hep-th]].

\bibitem{Hosomichi:2008ip}
K.~Hosomichi, K.~M.~Lee, S.~Lee, S.~Lee, J.~Park and P.~Yi,
``A Nonperturbative Test of M2-Brane Theory,''
JHEP {\bf 0811}, 058 (2008)
[arXiv:0809.1771 [hep-th]].

\bibitem{Hosomichi:2008jb}
  K.~Hosomichi, K.~M.~Lee, S.~Lee, S.~Lee and J.~Park,
  ``N=5,6 Superconformal Chern-Simons Theories and M2-branes on Orbifolds,''
  JHEP {\bf 0809}, 002 (2008)
  [arXiv:0806.4977 [hep-th]].

\bibitem{Benna:2008zy}
  M.~Benna, I.~Klebanov, T.~Klose and M.~Smedback,
  ``Superconformal Chern-Simons Theories and AdS${}_4$/CFT${}_3$ 
  Correspondence,''
  JHEP {\bf 0809}, 072 (2008)
  [arXiv:0806.1519 [hep-th]].

\bibitem{HLL} K.~Hosomichi, K.~Lee, S.~Lee, private communication.


\bibitem{SheikhJabbari:2009kr}
  M.~M.~Sheikh-Jabbari and J.~Simon,
  ``On Half-BPS States of the ABJM Theory,''
  arXiv:0904.4605 [hep-th].

\bibitem{Low:2009kv}
  A.~M.~Low,
  ``N=6 Membrane Worldvolume Superalgebra,''
  JHEP {\bf 0904} (2009) 105
  [arXiv:0903.0988 [hep-th]].

\bibitem{Grossman:1990it}
B.~Grossman,
``Hierarchy of soliton solutions to the gauged nonlinear Schrodinger equation
on the plane,''
Phys.\ Rev.\ Lett.\  {\bf 65}, 3230 (1990).

\bibitem{Dunne:1990qe}
G.~V.~Dunne, R.~Jackiw, S.~Y.~Pi and C.~A.~Trugenberger,
``Selfdual Chern-Simons solitons and two-dimensional nonlinear equations,''
Phys.\ Rev.\  D {\bf 43}, 1332 (1991)
[Erratum-ibid.\  D {\bf 45}, 3012 (1992)].

\bibitem{Taubes:1979tm}
C.~H.~Taubes,
``Arbitrary N: Vortex Solutions To The First Order Landau-Ginzburg
Equations,''
Commun.\ Math.\ Phys.\  {\bf 72}, 277 (1980).

\bibitem{GKKKN} G. Go, C. Kim, Y. Kim, O-K. Kwon, H. Nakajima,
in preparation.

\bibitem{Hanany:2003hp}
  A.~Hanany and D.~Tong,
  ``Vortices, instantons and branes,''
  JHEP {\bf 0307}, 037 (2003)
  [arXiv:hep-th/0306150].

\bibitem{Eto:2006pg}
  M.~Eto, Y.~Isozumi, M.~Nitta, K.~Ohashi and N.~Sakai,
  ``Solitons in the Higgs phase: The moduli matrix approach,''
  J.\ Phys.\ A  {\bf 39} (2006) R315
  [arXiv:hep-th/0602170].


\bibitem{Kim:1993mh}
  C.~Kim, C.~Lee, P.~Ko, B.~H.~Lee and H.~Min,
  ``Schrodinger fields on the plane with U(1)**N Chern-Simons interactions and
  generalized selfdual solitons,''
  Phys.\ Rev.\  D {\bf 48}, 1821 (1993)
  [arXiv:hep-th/9303131].

\end{thebibliography}
\end{document}